\documentclass[]{bmcart}

\usepackage[utf8]{inputenc} %unicode support
\usepackage{graphicx}

\startlocaldefs
\endlocaldefs

\begin{document}

%%% Start of article front matter
\begin{frontmatter}

\begin{fmbox}
\dochead{Research}


\title{Stochasticity in pandemic spread over the World Airline Network explained by local flight connections}

%%%%%%%%%%%%%%%%%%%%%%%%%%%%%%%%%%%%%%%%%%%%%%
%%                                          %%
%% Enter the authors here                   %%
%%                                          %%
%% Specify information, if available,       %%
%% in the form:                             %%
%%   <key>={<id1>,<id2>}                    %%
%%   <key>=                                 %%
%% Comment or delete the keys which are     %%
%% not used. Repeat \author command as much %%
%% as required.                             %%
%%                                          %%
%%%%%%%%%%%%%%%%%%%%%%%%%%%%%%%%%%%%%%%%%%%%%%

\author[
   addressref={aff1},                   % id's of addresses, e.g. {aff1,aff2}
   corref={aff1},                       % id of corresponding address, if any
%   noteref={n1},                        % id's of article notes, if any
   email={lawyer@mpi-inf.mpg.de}   % email address
]{\inits{GL}\fnm{Glenn} \snm{Lawyer}}

%%%%%%%%%%%%%%%%%%%%%%%%%%%%%%%%%%%%%%%%%%%%%%
%%                                          %%
%% Enter the authors' addresses here        %%
%%                                          %%
%% Repeat \address commands as much as      %%
%% required.                                %%
%%                                          %%
%%%%%%%%%%%%%%%%%%%%%%%%%%%%%%%%%%%%%%%%%%%%%%

\address[id=aff1]{%                           % unique id
  \orgname{Department of Computational Biology, Max Planck Institute for Informatics}, % university, etc
  \street{Campus E1 4},                     %
  %\postcode{}                                % post or zip code
  \city{Saarbrücken},                              % city
  \cny{DE}                                    % country
}
\address[id=aff2]{%
  \orgname{Marine Ecology Department, Institute of Marine Sciences Kiel},
  \street{D\"{u}sternbrooker Weg 20},
  \postcode{24105}
  \city{Kiel},
  \cny{Germany}
}

%%%%%%%%%%%%%%%%%%%%%%%%%%%%%%%%%%%%%%%%%%%%%%
%%                                          %%
%% Enter short notes here                   %%
%%                                          %%
%% Short notes will be after addresses      %%
%% on first page.                           %%
%%                                          %%
%%%%%%%%%%%%%%%%%%%%%%%%%%%%%%%%%%%%%%%%%%%%%%

\begin{artnotes}
%\note{Sample of title note}     % note to the article
%\note[id=n1]{Equal contributor} % note, connected to author
\end{artnotes}

\end{fmbox}% comment this for two column layout

%%%%%%%%%%%%%%%%%%%%%%%%%%%%%%%%%%%%%%%%%%%%%%
%%                                          %%
%% The Abstract begins here                 %%
%%                                          %%
%% Please refer to the Instructions for     %%
%% authors on http://www.biomedcentral.com  %%
%% and include the section headings         %%
%% accordingly for your article type.       %%
%%                                          %%
%%%%%%%%%%%%%%%%%%%%%%%%%%%%%%%%%%%%%%%%%%%%%%

\begin{abstractbox}
\begin{abstract} % abstract
\parttitle{Background} %if any
Massive growth in human mobility has dramatically increased the risk and rate of pandemic spread.
Macro-level descriptors of the topology of the  World Airline Network (WAN)  explains middle and late stage dynamics of pandemic spread mediated by this network, but necessarily regard early stage variation as stochastic.
We propose that much of early stage variation can be explained by appropriately characterizing the local topology surrounding the debut location of an outbreak.
\parttitle{Methods}
Based on a model of the WAN derived from public data, we measure for each airport the expected force of infection (AEF) which a pandemic originating at that airport would generate.
We  observe, for a subset of world airports, the minimum transmission rate at which a disease becomes pandemically competent at each airport. 
We also observe, for a larger subset,  the time until a pandemically competent outbreak achieves pandemic status given its debut location.
Observations are generated using  a highly sophisticated metapopulation reaction-diffusion simulator under a disease model known to well replicate the 2009 influenza pandemic.
The robustness of the AEF measure to model misspecification is examined by degrading the underlying model WAN.
\parttitle{Results} 
AEF powerfully explains pandemic risk, showing 
correlation of 0.90 to the transmission level needed to give a disease pandemic competence, and
correlation of 0.85 to the delay until an outbreak becomes a pandemic.
The AEF is robust to model misspecification. For 97\% of airports, removing 15\% of airports from the model changes their AEF metric by less than 1\%.

\parttitle{Conclusions} 
Appropriately summarizing the size, shape, and diversity of an airport's local neighborhood in the WAN accurately explains much of the macro-level stochasticity in pandemic outcomes.
\end{abstract}

%%%%%%%%%%%%%%%%%%%%%%%%%%%%%%%%%%%%%%%%%%%%%%
%%                                          %%
%% The keywords begin here                  %%
%%                                          %%
%% Put each keyword in separate \kwd{}.     %%
%%                                          %%
%%%%%%%%%%%%%%%%%%%%%%%%%%%%%%%%%%%%%%%%%%%%%%


% MSC classifications codes, if any
%\begin{keyword}[class=AMS]
%\kwd[Primary ]{}
%\kwd{}
%\kwd[; secondary ]{}
%\end{keyword}

\end{abstractbox}
%
%\end{fmbox}% uncomment this for twcolumn layout

\end{frontmatter}

%%%%%%%%%%%%%%%%%%%%%%%%%%%%%%%%%%%%%%%%%%%%%%
%%                                          %%
%% The Main Body begins here                %%
%%                                          %%
%% Please refer to the instructions for     %%
%% authors on:                              %%
%% http://www.biomedcentral.com/info/authors%%
%% and include the section headings         %%
%% accordingly for your article type.       %%
%%                                          %%
%% See the Results and Discussion section   %%
%% for details on how to create sub-sections%%
%%                                          %%
%% use \cite{...} to cite references        %%
%%  \cite{koon} and                         %%
%%  \cite{oreg,khar,zvai,xjon,schn,pond}    %%
%%  \nocite{smith,marg,hunn,advi,koha,mouse}%%
%%                                          %%
%%%%%%%%%%%%%%%%%%%%%%%%%%%%%%%%%%%%%%%%%%%%%%

%%%%%%%%%%%%%%%%%%%%%%%%% start of article main body
% <put your article body there>
%\input{Lawyer-manuscript}
%%%%%%%%%%%%%%%%
%% Background %%
%%
%

\section*{Background}
%\subsection*{lead-in}
The world airline network (WAN) has massively increased the speed and scope of human mobility. This boon for humanity has also created an efficient global transport network for infectious disease \cite{Tatem2006,Tatem2014}.
Pandemics can now occur more easily and more quickly than ever before.
The accelerating emergence of novel pathogens  exacerbates the situation \cite{Jones2008}.
Better understanding of global dispersal dynamics is a major challenge of our century \cite{Brockmann2013}.
Rapid assessment of an emerging outbreak's dissemination potential is critical to response planning \cite{Johansson2014}. 
We do not know where the next pandemic threat might emerge. Mexico was not a prime candidate for an influenza outbreak, nor West Africa for Ebola.
Preemptively mapping the pandemic influence of individual airports could contribute substantially to monitoring and response plans.

%\subsection*{Structure is main driver}
While exact relationships between the WAN and pandemic spread are difficult to model~\cite{Tatem2014}, simulation studies suggests that topological descriptors which describe epidemic outcomes on network models also have explanatory power for relationships between the topology of the WAN and pandemic spread \cite{Colizza2006,Colizza2007}.
 %, and later theoretical work supports these results \cite{Preciado2014_ieee}.
Observational studies of influenza \cite{Brockmann2013,Balcan2009}, malaria \cite{Huang2013}, and dengue fever \cite{Semenza2014} support this conclusion.
Given the topology of a network, 
the minimal disease transmission rate which allows epidemics is given by the inverse of the spectral radius of a network's adjacency matrix \cite{Heffernan2005}, 
and the typical outcome \cite{Newman2002} and time course \cite{Volz2008} of an epidemic follow a closed-form solution governed by the degree distribution of the network.
The WAN's topological structure is well characterized. 
It is a small-world, scale-free network with strong community structure, imposed partly by spatial constraints~\cite{Barrat2005}.
The majority of airports (70\%) serve as bridges which connect a densely interconnected core of 73 major transport hubs (2\%) to regional population centers and peripheral airports (28\%)  \cite{Verma2014}.
Nodes which connect communities can be distinct from high-degree nodes within communities~\cite{Guimera2005}.
Since the WAN is designed to optimize passenger flow, the network's temporal structure has little effect at time scales relevant for pandemic spread~\cite{Pan2011}.

%For example, when considering all possible networks with a given degree distribution, and all possible seeding points for a disease process,  the typical outcome~\cite{Newman2002} 
%Likewise, the effect of human mobility on the difference between invasion and non-invasion dynamics is crucially affected by the degree distribution of the mobility network~\cite{Colizza2007}.
%For example, epidemic diffusion is nearly instantaneous when transmission is mediated by a scale-free network~\cite{lloyd2001}, and 
%Spatial constraints push high degree nodes towards the barycenter of the network, while the need for short-range links counteracts a tendency to connect hubs while encouraging formation of cliques
% The core is densly interconnected, making it robust to failure of any one node; any one can easily be routed around. Failure of a bridge cuts off all associated periphery nodes. \emph{Implications: cannot shut core}

%\subsection*{Models}
Topological descriptors of epidemic dynamics, however, can only describe typical outcomes. They do not describe the structure of the variation around the typical outcome, which is dismissed as stochastic when mentioned at all.
Even within the constraints of a simple branching process model, empirical estimates of the probability of epidemic show substantial variation around the  analytically derived solution, see Figure \ref{fig:branching}.
 Actual outcomes of emergent infectious diseases are crucially shaped by chance events in the early phases of their emergence~\cite{Bauch2005}.
Clear understanding of how seed location influences global outcomes would substantially improve public health planning \cite{Johansson2014}.

%Rapid assessment of the potential for outbreaks to spread is critical to 
%At the macro level, this increases with $\beta$, the disease transmission rate, in a manner well described by branching process models.
%Likewise, the effect of human mobility on the invasion threshold depends on the degree distribution of the mobility network~\cite{Colizza2007}, with increased variance in degree distribution reducing the invasion threshold.
%These macro descriptor does not, however, offer insight into how the seed location influences the invasion probability. 

The development of sophisticated, parameter-rich epidemic simulators provides powerful tools for exploring relationships between seed location and epidemic outcomes \cite{Tizzoni2012}.
Common frameworks encompass  demographic and mobility characteristics via either metapopulation \cite{Balcan2009,Broeck2011}  or agent-based assumptions~\cite{Ajelli2011}.
Careful tuning of these models has produced results which well match the spread of the 2009 influenza epidemic \cite{Tizzoni2012,Ajelli2010}.
Yet the complex interactions between model structure, input parameters, and estimation methods makes interpretation of model-based results challenging~\cite{Bauch2005}, especially when attempting to generalized to future outbreaks for which epidemic parameters are fundamentally unknowable.
%Simple models which capture only relevant details provide more understanding~\cite{May2004}.
If, however, two radically different modeling approaches result in such high agreement both with each other and with reality~\cite{Ajelli2010} then the principal driver of outcomes should be expressible with a small parameter set \cite{Brockmann2013}.
Evidence suggests that simple probabilistic models incorporating local incidence, travel rates, and basic transmission parameters are sufficient to predict outcomes of complex metapopulation based simulations \cite{Johansson2012}.

%\subsection*{Measuring node influence}
Recent theoretical work suggests that the apparent stochasticity in the early phases of a network-mediated epidemic process can be explained by the expectation of the force of infection of epidemic processes seeded from that node \cite{Lawyer2015}.
The aim of this study is to evaluate if this finding generalizes to realistic scenarios of WAN-mediated pandemic disease spread.

%This is largely due to lack of appropriate measures; traditional centrality indicators, for example, while useful for identifying the most important nodes, are poor descriptors of the influence of the vast majority of nodes. 
%A sample of 2-300 includes the largest airports, the largest cities, the most-connected cities, and the most central cities is claimed to be sufficient to describe the dynamics of the global spread of influenza \cite{Bobashev2008}. 
%dynamics dominated by a small percentage of transport connections. \cite{Brockmann2013}

\section*{Methods}

\subsection*{Defining and measuring ExF on the WAN}
Our model of the WAN is based on the 2014 release of the Open Flights database \cite{openflights}.
We selected all airports serviced by regularly scheduled commercial flights, resulting in a list of 3,458 airports connected by 68,820 routes served by 171 different aircraft types. 
We simplify the network by replacing multiple routes between airports by a single edge whose weight is the sum of the available seats on all routes connecting the two airports, under the assumption that the aircraft type reflects the airline's best judgment of the importance of the route.
Aircraft seating capacity was estimated based on aircraft descriptions on worldtrading.net and airliners.net, using airlinecodes.co.uk to translate the IATA aircraft codes into aircraft type.

The expected force of a network node is defined as the expectation of the force of infection generated by an epidemic process seeded from the node into an otherwise fully susceptible network, after two transmission events and no recovery~\cite{Lawyer2015}. 
The force of infection in a network is directly proportionate to the current number of infected-susceptible edges or in a weighted network, the sum of all such edge weights. 
Its expectation  after two transmission is given by the entropy of the distribution of this sum over all possible ways the first two transmission could occur.
Applied to the WAN, 
\[
AEF(i)=- \sum_{j=1}^{J} \bar{d_j} \log(\bar{d_j})
\] % \end{equation}
where 
$AEF(i)$ is \textbf{A}irport i's \textbf{E}xpected \textbf{F}orce, 
$J$ enumerates all possible ways to observe two transmissions seeded from $i$, 
$d_j$ is the weighted degree of the $j^{th}$ transmission pattern multiplied by the probability that this pattern is observed given $J$, and 
$\bar{d_j}=d_j /\sum_{kj=1}^J d_k$ is the normalization of $d_j$.
We here further normalize AEF values to the range $[0,100]$.
%Do I put in the

Epidemic outcomes are generated using the GLEaMvis simulator \cite{Balcan2009,Broeck2011}.
GLEaMvis integrates real-world global population and mobility data with an individual based stochastic mathematical model of the infection dynamics to produce realistic simulations of the global spread of infectious diseases.
Our basic experimental setup is to simulate the same disease model over a range of seed cities.
The structure and parameters of the disease models are based on those which match the 2009 Influenza pandemic as reported in \cite{Balcan2009} and validated in \cite{Tizzoni2012}, 
specifically, a \textbf{S}usceptible-\textbf{E}xposed-\textbf{I}nfected-\textbf{R}ecoverd (SEIR) model with transmission rate $\beta$ specified below, latency rate $\epsilon=1/1.1$ and  recovery rate $\mu = 1/2.5$. 
Rates are expressed in units of days.
The initial population distribution is 10\% of the seed city infected and the remainder of the (world) population susceptible.
Seasonality effects are not included, since their influence varies both by time and geographic latitude, masking variability attributable to seed location.
GLEaMvis divides the world into sixteen regions. 
An outbreak is declared a pandemic on the day prevalence in at least three regions is greater than one per 100,000 inhabitants.
The pattern the results is invariant to thresholds in the range $[0.1, 100]$ per 100,000 inhabitants and to replacing the ``three regions'' criteria with ``100 cities.''
Results for each airport are reported in terms of the median over 20 runs (the maximum number supported by the public GLEaMvis client).
If the threshold is not passed after 365 days (the maximum length supported by the public GLEaMvis client), we declare that no pandemic occurred.

%This model is a radical simplification of the forces shaping global disease spread.  Most glaring is that the WAN is only one aspect of human mobility. Second, our model ignores obvious links between airports; for example, we treat London Heathrow and London Gatwick as independent entities though both service the same population. Third, our model is a single  time-point medium resolution snapshot of a dynamic system. This simplification is intentional and in keeping with our hypothesis, that a small parameter set can explain much of the variance in a much larger, more complex system.

\subsection*{Defining and measuring epidemic stochasticity}
For an outbreak to become a pandemic, 
its basic reproductive number $R_0$ must surpass the  basic epidemic threshold  $R_0>1$ needed to establish a disease in a local population by a sufficient amount to also overcome finite subpopulation size effects and diffusion rates to neighboring populations.
A branching process approximation suggests that invasion thresholds in metapopulation models depend on the outbreak's $R_0$ value, the variance of the network's degree distribution, and the mobility rate between subpopulations \cite{Colizza2007}. 
The GLEaMviz model specifies the last two values, reducing invasion thresholds to a function of $R_0$.
However, as shown in Figure 1, even a pure branching process shows substantial variability around the theoretical probability of achieving a large outbreak. For pandemics mediated by the WAN, the question of interest is how the invasion threshold varies for different airports.
We empirically observe invasion thresholds on the WAN as follows.
Ten seed airports are selected, one from each decile of the range of 
AEF values, see Table \ref{tab:seeds}.
The basic reproductive number is defined as $R_0 = \beta/\mu$, the transmission to recovery ratio. Keeping $\mu$ fixed, we vary $\beta$ over the range [0.4, 0.5], and observe which seeds  trigger a pandemic at each value under the simulation framework described above.
For $\beta< 0.4$, no simulations reached pandemic status, and for $\beta>0.45$, all simulations resulted in a pandemic. 
%We continue running simulations for $\beta>0.45$  to observe how the the date of pandemic varies as $beta$ increases for the different airports.
%Testing pandemic risk on real-world data is difficult in that it would require collecting epidemiological data from pandemics which never occurred.

Often, diseases of concern are known to be competent of invading the network. 
Here, the outcome of interest is not if a pandemic occurs, but rather how long until an outbreak reaches pandemic status.
We measure relationships between AEF and time to pandemic status  as follows. 
One hundred world airports were chosen such that they evenly cover the range of measured AEF values.
The simple SEIR model used previously is extended to the full model used by the GLEaMvis group to replicate the 2009 pandemic \cite{Balcan2009}. This estimates transmission rate as  $\beta=0.8383$, well above the invasion threshold of $\beta \approx 0.45$ determined empirically above.
Further, the infected compartment of the SEIR model is divided into three categories: asymptomatic, symptomatic travelers, symptomatic non-travelers. These categories affect the mobility model, and non-symptomatic individuals have reduced transmissibility. 
For each seed location, we observe both the number of days until pandemic status is reached and the number of days until peak global incidence.
Both outcomes are highly correlated, since once pandemic status is achieved further disease development is determined by network topology.
The purpose of measuring peak global incidence is that this measure is unambiguous, while any definition of ``first day of pandemic status'' is somewhat arbitrary.
A Shapiro-Wilks test of the observed times to peak global incidence suggests that this data is approximately normally distributed ($p=0.69$ under the null hypothesis that the data is normally distributed),
while the distribution of observations of first day of pandemic status is right-skewed ($p=0.04$).

Relationships between outcomes and AEF are measured by Pearson correlation. 
We additionally test correlations to weighted and unweighted versions of each airport's betweenness, degree, and eigenvalue centralities, and also to Verma et al's t-core, a variant of the k-core which counts triangles \cite{Verma2014}.

%\subsection*{Real Epidemic outcomes}
%\emph{Need data on 2009 inf and SARS; also, what do I measure/observe?}

\subsection*{Robustness of AEF to sampling error}
The robustness of AEF values is examined by observing their relative change while progressively degrading the model WAN from which they are derived.
The network is degraded by removing from one to 15 percent of U.S. airports from the network along with their associated edges.
Community-based analysis of the WAN suggest that US airports form one large community \cite{Verma2014,Guimera2005}.
The AEF of all remaining world airports is computed.
Three different random removal schemes were used: uniform over all airports, selection weighted by airport degree, selection weighted by AEF.
The resulting AEF values are compared with the original AEF values. 
We record the number of airports whose degraded AEF departs from its original AEF by more than 1\% and by more than 5\%.
Reported results are the average over ten runs, and show the amount of degradation for both U.S. and non-U.S. airports.

\section*{Results}
The AEF of the seed location is strongly predictive of an outbreak's invasive threshold as shown in Figure \ref{fig:fitbeta} and Table \ref{tab:seeds}. The correlation between AEF and the minimal observed transmission rate at which it first became pandemically competent was 0.90 (95\% confidence interval: 0.98,0.62). 
Tokyo was a notable outlier, achieving pandemic competence earlier than predicted from its AEF value.

AEF was also strongly correlated with the delay until an outbreak became a pandemic. Correlation was $0.84 \pm 5.8$
to the day pandemic status was achieved, and $0.85 \pm 5.6$
to the day of peak global incidence, see Figure \ref{fig:ttp}.
AEF is significantly and more strongly correlated to either epidemic outcome than any of the comparison network centrality measures, see Table \ref{tab:cor} and Figure \ref{fig:mcomp}.

The  AEF proved robust to incomplete sampling.
Degradation was most severe when airports were preferentially removed based on degree. 
Still, only three percent of non-U.S. airports showed more than 1\% change in their computed ExF values when applying this scheme at the highest noise level.
Even within the United States, only 22\% of AEF values changed by more than 5\%.
See Figure \ref{fig:deg}.

\section*{Discussion}

%These results clearly demonstrate that the early stages of pandemic spread on the world airline network is are not stochastic variables, but can be determined from features of the seeding airport.
In all cases, AEF explains much of the variation in epidemic outcomes, suggesting that the early development of a pandemic is not stochastic, but rather strongly structured by the local connectivity of the seed location.
The ability of the AEF to summarize this connectivity contributes substantially to our understanding of the role of individual airports in pandemic diffusion.
These results are in harmony with other recent work claiming that relative arrival times of WAN-mediated pandemics are independent of disease-specific parameters~\cite{Brockmann2013} and that a simple branching process model describes early developments as well as complex metapopulation simulations \cite{Johansson2012}.

%\subsection*{outliers}
Degradation of the network had, in general, limited effect on airport AEF values. 
Wrong information regarding a specific node could, however, produce a misleading AEF value for that airport.
Epidemics seeded from airport PBJ (Paama Island, Vanuatu) took longer than expected to achieve pandemic status. This airport is probably mischaracterized in the Open Flights database, as flights to this simple grass strip are not shown on the Vanuatu airlines online booking system (\texttt{http://www.airvanuatu.com/}, last visited 23 March 2015).
In the opposite direction, Narita Airport (NRT, Tokyo, Japan) showed significantly greater pandemic risk than predicted by its AEF. This could be due to Japan's intense population density combined with high local mobility, factors captured in the GLEaMvis simulator but not the Open Flights database.

Two outliers highlight a structural blind spot of the AEF metric. Epidemics seeded from airports ZRJ (Round Lake, Canada) and PVH (Porto Velho, Brazil) took longer than expected to achieve pandemic status.
ZRJ is part of a small but locally dense community of airports serving first nation communities in Canada. This community has limited connectivity to the rest of the WAN, and ZRJ is three flights distant from any airport outside this community (Winnipeg's James Armstrong Richardson Airport YWG, Chicago Midway MDW, Toronto Pearson YYZ).
Likewise, PVH is two flights from any of Brazil's international transport hubs.
The AEF is here derived from an airport's two-hop neighborhood, meaning for certain airports it is unaware of these network community boundaries. 
This limitation could perhaps be overcome by instead computing AEF based on a three-hop neighborhood. Given, however, that the WAN's effective diameter is four hops, and the general good performance of the AEF, it is not clear that such an extension would substantially improve results globally.

%. This is an example of the AEF being mislead by a locally dense community which is otherwise only loosely connected to the rest of the network. Round Lake--Sioux Lookout-- Thunder Bay, and then, if lucky, to Chicago, Toronto, or Winnipeg.
% Same is true, but to a lesser extent, for PVH.

Airport expected force summarizes the size, density, and diversity of each airport's neighborhood in the WAN.
It combines features of degree, neighbor degree, and betweenness centrality in a statistically coherent manner.
Airport degree is not a good descriptor of pandemic outcomes, since it does not account for a neighbor's onward connectivity.
Guimera et al noted that high degree does not well correlate to high centrality \cite{Guimera2005}. Nor does low degree correlate to an airport's connection to the wider network, as illustrated by comparing Sweden's Link{\"o}ping City Airport (LPI) to Alaska's Huslia Airport (HSL).  HSL has four outbound routes which connect to other rural Alaskan airports. LPI has only one outbound route, which connects to Amsterdam Schipol.
Verma et al propose instead characterizing airports based on the number of network triangles they take part in, the t-core \cite{Verma2014}.
Plotting airport t-core against epidemic outcomes shows that its ability to explain epidemic outcomes is a result of its ability to successfully segment the WAN into core and periphery, see Figure \ref{fig:mcomp}.
Thus  t-core and AEF capture complementary aspects of an airport's role in the WAN.

%\subsection*{Extensions}
%The AEF as defined here 
The applicability of the AEF could be extended by modifying it to allow for varying transmissibility at individual airports. 
Such an extension would allow it to express differences in i.e.  competent vector species populations or health care system readiness at different world locations.
Since the AEF is the expectation of the force of infection, such an extension merely requires modifying the calculation of each transmission pattern's force of infection along with the probability of that specific pattern occurring.
Both criteria can be met by adjusting edge weights in the underlying network model, implying that this extension could be implemented using the same framework as outlined in the current work.
It would also be interesting to apply the expected force framework to disease spread through the world shipping network, a major transport system for several vector born pathogens along with their vector species \cite{Tatem2006}.
The approach could also be tested on more local transmission network models, such as contacts in a hospital ward \cite{Machens2013} or city-wide mobility data acquired from i.e. mobile phones \cite{Tizzoni2014,Deville2014}.

\section*{Conclusion}

An outbreak's debut location is highly influential in its ability to become a pandemic threat. The AEF metric succinctly captures this influence, and can help inform monitoring and response strategies.

These investigations pave the way for the development of simple, robust models capable of informing preparedness planning and policy directives.

%%%%%%%%%%%%%%%%%%%%%%%%%%%%%%%%%%%%%%%%%%%%%%
%%                                          %%
%% Backmatter begins here                   %%
%%                                          %%
%%%%%%%%%%%%%%%%%%%%%%%%%%%%%%%%%%%%%%%%%%%%%%

\begin{backmatter}

\section*{Competing interests}
  The Max Planck Society has filed for a patent on the use of the expected force metric to assess spreading risk on the world airline network.
\section*{Author's contributions}
 GL concieved and carried out the experiments and wrote the manuscript.
\section*{Acknowledgements}
 We thank the GLEaMvis team for providing public access to their simulator with the only requirement being appropriate citation. Trivik Verma provided measures of airport t-core.

%%  After submission of the .TEX file,                     %%
%%  you will be prompted to submit your .BBL file.         %%

% if your bibliography is in bibtex format, use those commands:
\bibliographystyle{bmc-mathphys} % Style BST file


% or include bibliography directly:
% \begin{thebibliography}
% \bibitem{b1}
% \end{thebibliography}
\pagebreak
%%%%%%%%%%%%%%%%%%%%%%%%%%%%%%%%%%%
%%                               %%
%% Figures                       %%
%%                               %%
%% NB: this is for captions and  %%
%% Titles. All graphics must be  %%
%% submitted separately and NOT  %%
%% included in the Tex document  %%
%%                               %%
%%%%%%%%%%%%%%%%%%%%%%%%%%%%%%%%%%%

%%
%% Do not use \listoffigures as most will included as separate files

\section*{Figures}
%  \begin{figure}[h!]
%  \caption{\csentence{Sample figure title.}
%      A short description of the figure content should go here.}
%\end{figure}

\begin{figure}[h!]
\includegraphics[width=0.9\textwidth]{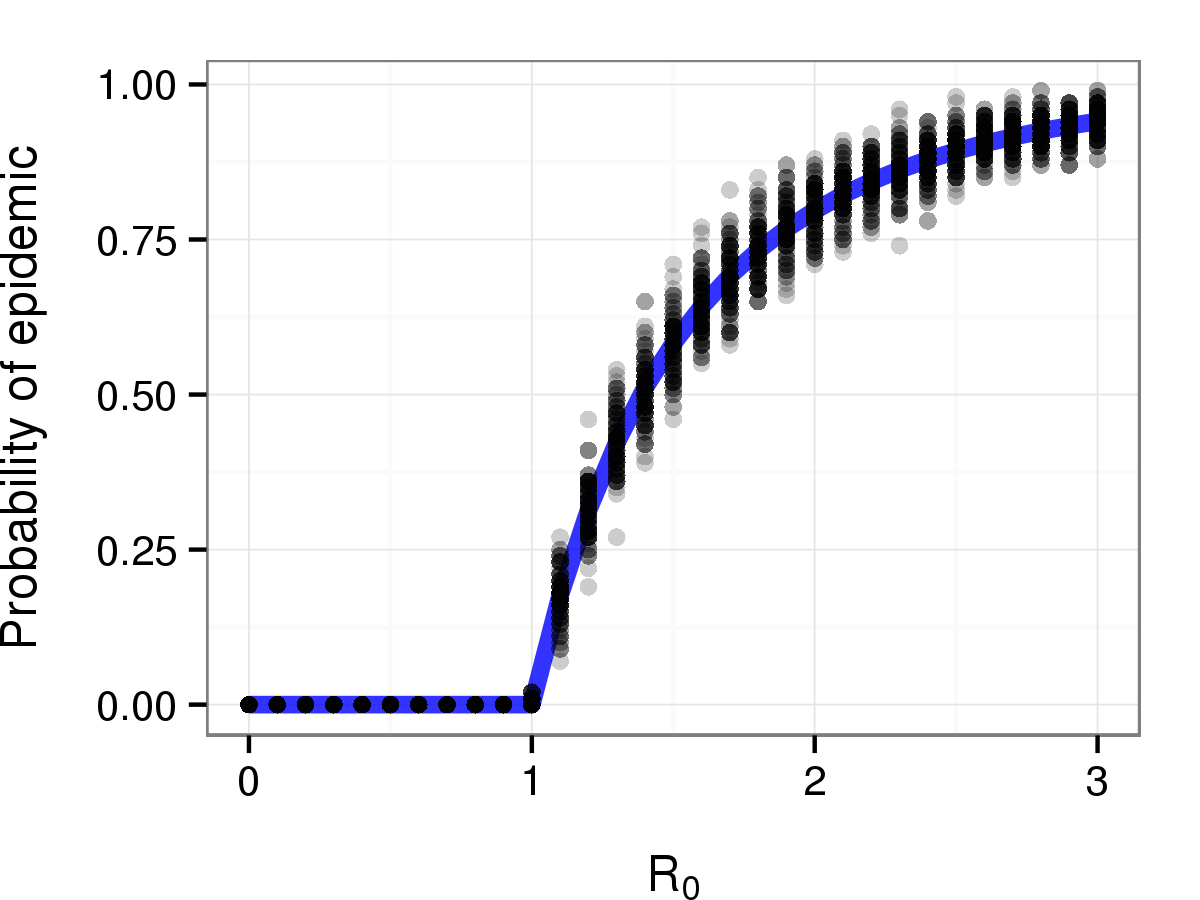}
\caption{ \csentence{Stochastic variation around the probability of a major outbreak under a simple branching process model.} In a discrete time Reed-Frost branching process with finite population, the probability of a major outbreak is the smallest solution to $x = e^{-R_0(1-x)}$, shown above as the solid blue line, where $R_0$ is the base reproductive number of the disease process. The black dots show empirically observed probabilities from simulations of the same model. Each dot is the observed fraction of major outbreaks out of 100 simulated outbreaks for a given value of $R_0$. For each value of $R_0$, 100 dots are generated.}
\label{fig:branching}
\end{figure}

%\begin{figure}[h!]
%\includegraphics[width=0.9\textwidth]{panrisk.png}
%\caption{Pandemic risk is well explained by airport expected force (right panels). The distribution (left panels) is stochastic, and strongly bimodal. Upper is unweighted, lower is weighted version.}
%\label{fig:panrisk}
% made by sisspreadrisk.R
%\end{figure}

\begin{figure}[h!]
\includegraphics[width=0.9\textwidth]{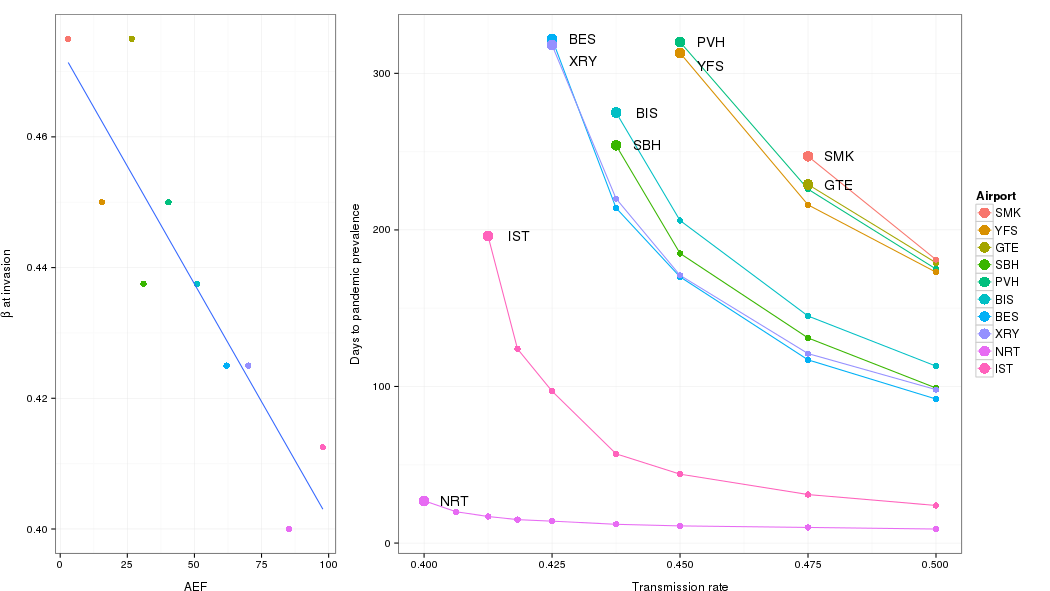}
\caption{\csentence{AEF and invasion threshold.}
Higher AEF is associated with a lower transmission rate needed to trigger pandemics, as well as shorter delay until the outbreak reaches pandemic status. Large dots mark the lowest transmission rate for which a pandemic occurred, for each city. Cities are listed in order of increasing AEF. See also Table \ref{tab:seeds}.}
\label{fig:fitbeta}
% made by gleam/fitbeta.R
\end{figure}

\begin{figure}[h!]
\includegraphics[width=0.9\textwidth]{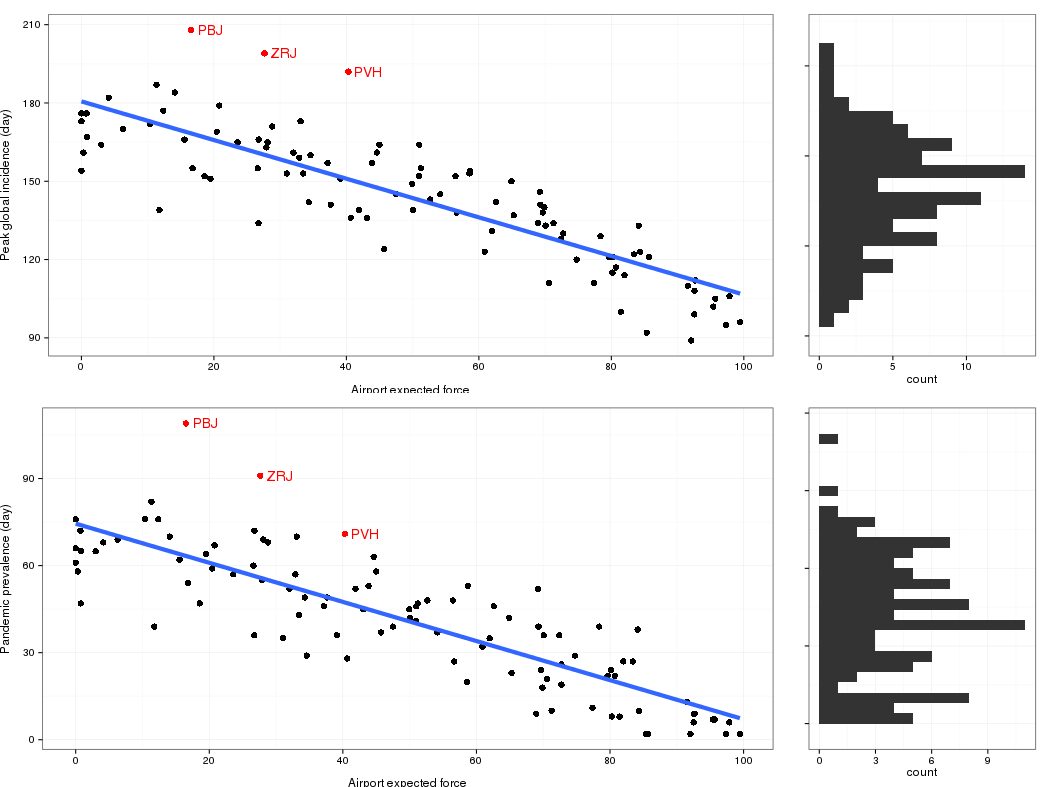}
\caption{\csentence{AEF and time to pandemic.}
Airport expected force has strong correlation to the rate at which a simulated epidemic becomes pandemic, an outcome which otherwise appears stochastic, as shown in the histograms in the right column.
The top panel shows days to peak global incidence, the bottom days until the disease achieves pandemic prevelance.
Outliers in the top plot are PBJ (Paama Island, Vanuatu), ZRJ (Round Lake, Canada), and PVH (Porto Velho, Brazil). }
% made by gleam/testPredPower.R
\label{fig:ttp}
\end{figure}

\begin{figure}[h!]
\includegraphics[width=0.9\textwidth]{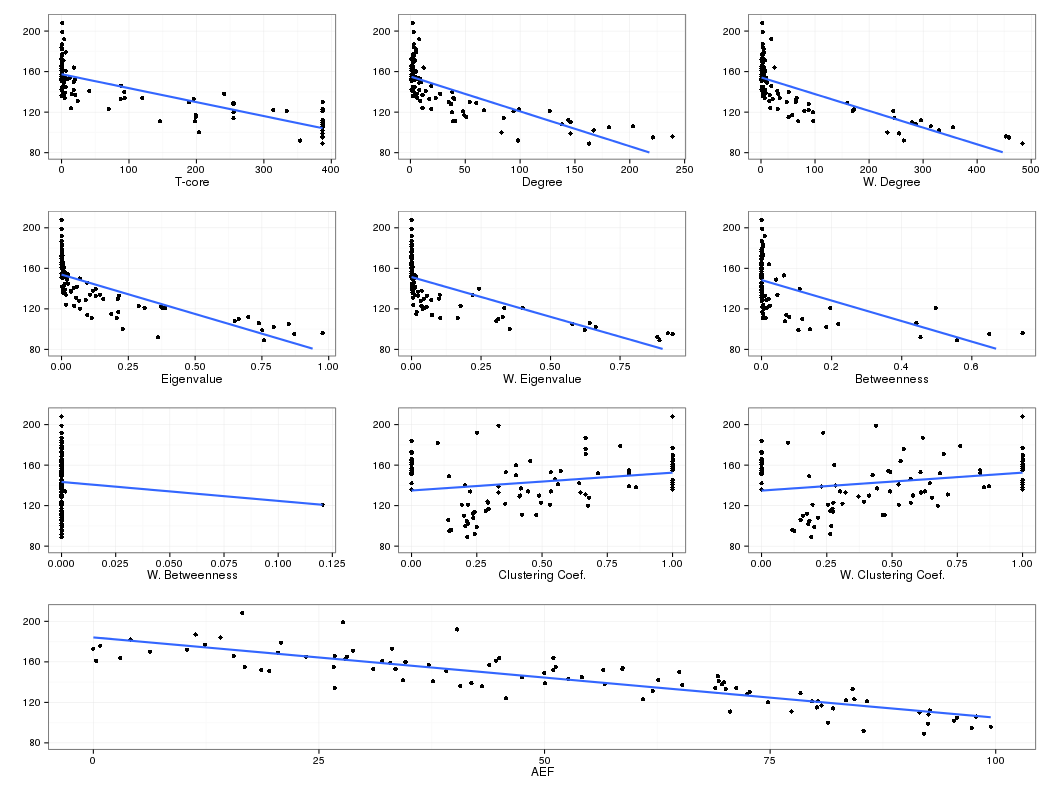}
\caption{\csentence{Centrality measures and time to peak global incidence}
Common network centrality measures do not well explain variation in the timing of peak global incidence.
The abbreviation ``W'' indicates the weighted version of the measure.}
% made by tcore/tcorecomp.R
\label{fig:mcomp}
\end{figure}

\begin{figure}[h!]
\includegraphics[width=0.9\textwidth]{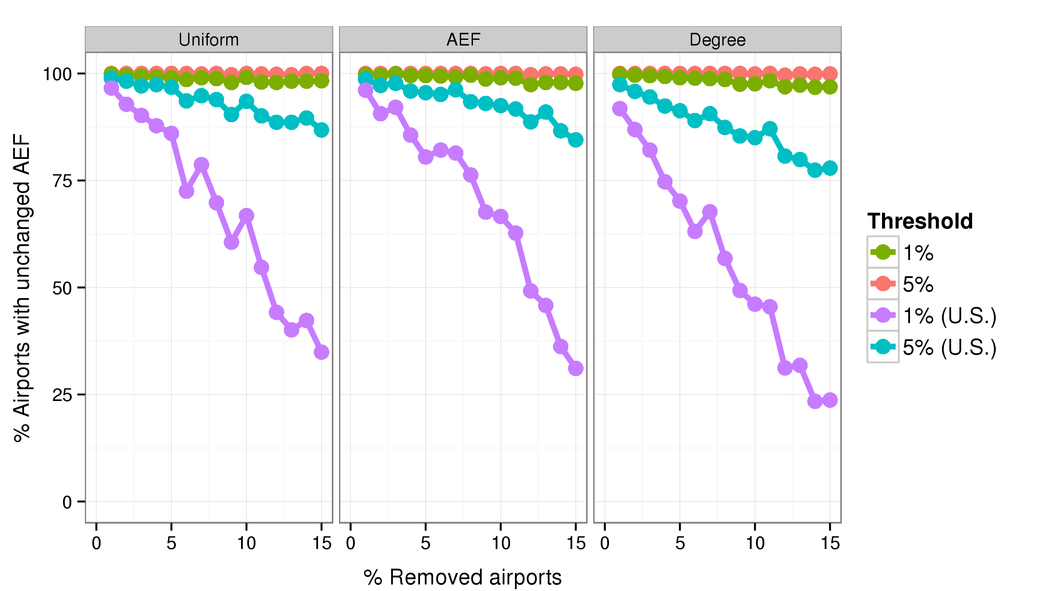}
\caption{\csentence{Robustness of AEF values to model degradation.}
For non-US airports, almost all AEF values are unaffected (defined as less than 1\%/5\% change) as US airports are removed from the model. 
While many US airports are affected at the 1\% level, few show more than 5\% change in AEF.
The network is degraded by removing airports with selection probabilities weighted uniformly, by AEF, and by degree.}
a\label{fig:deg}
% made by report/degrade.R
\end{figure}

%%%%%%%%%%%%%%%%%%%%%%%%%%%%%%%%%%%
%%                               %%
%% Tables                        %%
%%                               %%
%%%%%%%%%%%%%%%%%%%%%%%%%%%%%%%%%%%

%% Use of \listoftables is discouraged.
%%
\section*{Tables}

\begin{table}[h!]
\caption{Seed locations. The following airports were selected as seed locations for testing relationships between AEF and invasion risk.
The table additionally reports the number of days for an outbreak to reach pandemic status (``Pand.'') at the minimal observed transmission rate ($\beta$) for which a pandemic occured, along with each airport's t-core, (un)weighted degree, and (un)weighted eigenvalue centralities.
%Abbreviations: IATA - the IATA code for the airport. toff - days until pandemic prevalence. beta - lowest transmission rate which resulted in pandemic prevalence. AEF - airport expected force. T-core - airport t-core. 
}
\label{tab:seeds}
% made by gleam/fitBeta.R
\begin{tabular}{rlllrrrrrrrr}
  \hline
 & IATA & City & Country & Pand. & $\beta$ & AEF & t-core & Deg. & w. Deg. & Eigen. & w. Eigen. \\ 
  \hline
8 & SMK & St. Michael & USA  & 286 & 0.45 & 3 &   1 &   2 &   4 & 0.00 & 0.00 \\ 
  10 & YFS & Fort Simpson & Canada & 287 & 0.45 & 16 &   0 &   1 &   2 & 0.00 & 0.00 \\ 
  3 & GTE & Groote Eylandt & Australia & 300 & 0.45 & 27 &   3 &   3 &   4 & 0.00 & 0.00 \\ 
  7 & SBH & Gustavia & France & 346 & 0.42 & 31 &   7 &   6 &   9 & 0.01 & 0.00 \\ 
  6 & PVH & Porto Velho & Brazil & 295 & 0.45 & 40 &   4 &   8 &  19 & 0.00 & 0.00 \\ 
  2 & BIS & Bismarck & USA & 360 & 0.42 & 51 &   5 &   4 &   5 & 0.01 & 0.00 \\ 
  1 & BES & Brest & France & 284 & 0.42 & 62 &  24 &   9 &  15 & 0.05 & 0.00 \\ 
  9 & XRY & Jerez & Spain & 294 & 0.42 & 70 &  88 &  17 &  22 & 0.13 & 0.02 \\ 
  5 & NRT & Tokyo & Japan &  22 & 0.40 & 85 & 354 &  98 & 264 & 0.36 & 0.88 \\ 
  4 & IST & Istanbul & Turkey & 172 & 0.41 & 98 & 387 & 203 & 314 & 0.74 & 0.64 \\ 
   \hline
\end{tabular}
\end{table}

\begin{table}[h!]
\caption{Correlation and 95\% confidence interval between a suite of network centrality measures (rows) and days until both pandemic status and peak global incidence. The abbreviation ``W'' indicates the weighted form of the measure.}
\label{tab:cor}
% generated by tcore/tcorecomp.R
\begin{tabular}{llrrrr}
  \hline
 Measure & Pandemic & c.i. & Peak & c.i. \\ 
  \hline 
 AEF & -0.84        & $\pm 0.06$ & -0.85 & $\pm 0.06$ \\ 
 t-core & -0.77     & $\pm 0.09$ & -0.79 & $\pm 0.08$ \\ 
 Degree & -0.72     & $\pm 0.10$ & -0.75 & $\pm 0.09$ \\ 
 W degree & -0.69   & $\pm 0.11$ & -0.75 & $\pm 0.09$ \\ 
 Eigenvalue & -0.70 & $\pm 0.11$ & -0.74 & $\pm 0.09$ \\ 
 W eigenvalue & -0.66   & $\pm 0.12$ & -0.69 & $\pm 0.11$ \\ 
 Betweenness & -0.54& $\pm 0.14$ & -0.56 & $\pm 0.14$ \\ 
 W betweenness& -0.17   & $\pm 0.20$ & -0.09 & $\pm 0.20$ \\ 
 Clustering coef.  & 0.28 & $\pm 0.19$ & 0.24 & $\pm 0.19$ \\ 
W Clust. coef. & 0.27   & $\pm 0.19$ & 0.24 & $\pm 0.19$ \\ 
   \hline
\end{tabular}
\end{table}

\end{backmatter} %% should go after additional files
\end{document}